\documentstyle[12pt]{article}
\def\d{\delta}\def\e{\epsilon}

\newcommand{\p}[1]{(\ref{#1})}
\begin{document}
\newcommand{\eqn}[1]{eq.(\ref{#1})}
 
\renewcommand{\section}[1]{\addtocounter{section}{1}
\vspace{5mm} \par \noindent
  {\bf \thesection . #1}\setcounter{subsection}{0}
  \par
   \vspace{2mm} } 
\newcommand{\sectionsub}[1]{\addtocounter{section}{1}
\vspace{5mm} \par \noindent
  {\bf \thesection . #1}\setcounter{subsection}{0}\par}
\renewcommand{\subsection}[1]{\addtocounter{subsection}{1}
\vspace{2.5mm}\par\noindent {\em \thesubsection . #1}\par
 \vspace{0.5mm} }
\renewcommand{\thebibliography}[1]{ {\vspace{5mm}\par \noindent{\bf 
References}\par \vspace{2mm}} 
\list
 {\arabic{enumi}.}{\settowidth\labelwidth{[#1]}\leftmargin\labelwidth
 \advance\leftmargin\labelsep\addtolength{\topsep}{-4em}
 \usecounter{enumi}}
 \def\newblock{\hskip .11em plus .33em minus .07em}
 \sloppy\clubpenalty4000\widowpenalty4000
 \sfcode`\.=1000\relax \setlength{\itemsep}{-0.4em} }

\begin{flushright}
HUB EP-98/43\\
DFPD 98/TH/36\\
hep-th/9807133
\end{flushright}

\vspace{1cm} 
\begin{center}
{\bf Harmonics, Notophs and Chiral Bosons}

\bigskip
Paolo Pasti${}^*$, Dmitri~Sorokin ${}^*,{}^{**}$\footnote{
Alexander von Humboldt fellow.\\
On leave from Kharkov Institute of
Physics and Technology, Kharkov, 310108, Ukraine.} 
and Mario Tonin$^*$

\bigskip
${}^*$ Universit\`a Degli Studi Di Padova,
Dipartimento Di Fisica ``Galileo Galilei''\\
ed INFN, Sezione Di Padova\\
Via F. Marzolo, 8, 35131 Padova, Italia\\

\bigskip
${}^{**}$ Humboldt-Universit\"at zu Berlin\\
Mathematisch-Naturwissenshaftliche Fakultat\\
Institut f\"ur Physik\\
Invalidenstrasse 110, D-10115 Berlin, Germany
\end{center}

\centerline{ABSTRACT}

\vspace{- 4 mm}  
\begin{quote}\small
A way of covariantizing duality symmetric actions is described.
\end{quote}
  \par
   \vspace{2mm} 

\noindent 
The presence of self--dual fields or, in more general case, 
duality--symmetric fields in 
field--theoretical and string models reflects their duality properties 
whose extreme importance for understanding a full quantum theory has 
been appreciated during an impetuous development of the duality field
happened during last few years. The knowledge of duality--symmetric
effective actions is useful for carrying out more systematic study
of the classical and quantum properties of the theory,
and in this memorial contribution we would like to demonstrate how 
fruitful physical ideas and mathematical 
techniques which Victor Isakovich Ogievetsky and his colleagues have 
developed helped us to construct a covariant Lagrangian formulation 
applicable to all known models with duality--symmetric fields in 
space--time of Lorentz signature.

The problem of constructing and studying models described by 
duality--invariant actions has a rather long history. It goes back to
time when Poincare and later on Dirac noticed electric--magnetic duality 
symmetry of the free Maxwell equations, and, 
Dirac assumed the existence of magnetically charged particles 
(monopoles and dyons) \cite{dirac}
admitting the duality symmetry to be also held 
for the Maxwell equations in the presence of charged sources. 
To describe monopoles and dyons on an 
equal footing with electrically charged particles one should have
a duality--symmetric form of the Maxwell action. This problem was 
studied (among others) by Schwinger and Zwanziger, and in 1971 Zwanziger 
proposed a duality--symmetric action for Maxwell fields interacting with
dyonic sources \cite{z}. An alternative duality--symmetric Maxwell action
was proposed by Deser and Teitelboim in 1976 \cite{d}. The two actions, 
which proved to be dual to each other \cite{mps}, are not 
manifestly Lorentz--invariant. This feature 
turned out to be a general one.
Duality and space--time symmetries hardly coexist in one and the same 
action.

Later on this problem arose in multidimensional supergravity theories in
space--time of a dimension $D=4p+2$ where one would like to 
know how to construct
an action for self--dual tensor fields (chiral bosons) 
which are present in some versions
of supergravity and in the heterotic string. One of the ways of solving this
problem is to sacrifice Lorentz covariance in favour of duality 
symmetry. A non--covariant action for $D=2$ chiral bosons was 
constructed by Floreanini and Jackiw \cite{j}, and Henneaux and 
Teitelboim \cite{ht} proposed non--covariant actions for self--dual fields in
higher dimensional $D=4p+2$ space--time. 
In a context of modern aspects of duality Tseytlin \cite{ts} considered a 
duality--symmetric action for a string. 
Finally, Schwarz and Sen \cite{ss} constructed 
non--covariant duality--symmetric actions for dual tensor fields in any 
space--time dimension.

There have also been developed covariant approachs to the construction
of duality--symmetric actions. These use auxiliary fields.
The first covariant Lagrangian formulation of chiral bosons was proposed
by Siegel \cite{si} and its modification was considered by Kavalov and
Mkrtchyan \cite{km} in application to D=6 and D=10 chiral supergravities.
Another covariant approach is based on the use of an infinite number of 
auxiliary fields \cite{inf,ber}. It might be interesting that an effective 
self--dual action of this kind was extracted 
from a type IIB string field theory \cite{ber}.  

The third formulation was proposed in \cite{pst}. 
In its minimal version only one scalar auxiliary field is used 
to ensure space--time covariance of duality--symmetric actions. 
This approach turned out to be the most appropriate for the construction 
of the worldvolume action for the M-theory five--brane \cite{pst1}, 
duality--symmetric D=11 supergravity \cite{bbs} and D=10 IIB supergravity
\cite{alst}.

Below we will use Maxwell theory to demonstrate how this third approach 
was developed with promptings provided by works of V. I. Ogievetsky. 

It is well known that the standard action for  
a free Maxwell field
is not invariant under duality transformations of its electric and 
magnetic strength vectors.
To have a duality symmetry at the level of action one should double the 
number of gauge fields ($A^\alpha_m$, $\alpha=1,2$, $m$=0,1,2,3) \cite{z,d,ss} 
and construct an action in such a way that one of 
the gauge fields becomes an auxiliary field upon solving equations of 
motion.  The duality symmetric action of 
refs. \cite{d,ss} can be written in the following form:
\begin{equation}\label{ss}
S=\int d^4x(-{1\over 8}F^\alpha_{mn}F^{mn\alpha}
+{1\over 4}{\cal F}^\alpha_{0i}{\cal F}^{\alpha}_{i0}),\qquad (i=1,2,3)
\end{equation}
where 
\begin{equation}\label{sd}
{\cal F}^\alpha_{mn}={\cal L}^{\alpha\beta}F^\beta_{mn}-{1\over 
2}\epsilon_{mnlp}F^{lp\alpha}={1\over 2}\epsilon_{mnlp}
{\cal F}^{lp\beta}{\cal L}^{\alpha\beta},
\end{equation}
(${\cal L}^{12}=-{\cal L}^{21}=1$) is the self--dual combination of the 
field strengths.

The Zwanziger action \cite{z} differs from  \p{ss} by the sign
in front of the second term and in that, instead of the time--coordinate 
index, one of the spatial indices is separated in the analogous term
of the Zwanziger action.

Duality symmetry is a discrete subgroup of $SO(2)$ rotations of 
$A^\alpha_m$
$(A^\alpha_m~\rightarrow~{\cal L}^{\alpha\beta}A^\beta_m)$.

Note that because of the self--duality 
property \p{sd} ${\cal F}^\alpha_{mn}{\cal F}^{\alpha mn}\equiv 0$, 
and the best thing which one can do is to take the square of
only a part of the components of ${\cal F}^\alpha_{mn}$ for the 
construction of the second term of the action \p{ss}, 
and this breaks manifest Lorentz invariance. 

Here is a place to explain why the signature of space--time is important
for the possibility of applying the Lagrangian approach considered to 
the description of chiral bosons. It is crucial for this 
approach that the ``square" of a self--dual tensor is zero, which
holds, for instance, in $D=2p+2$ spaces of a Lorentz signature. Then 
taking the square of an appropriate part of the components of the
self--dual tensor (as in \p{ss}) one gets the desirable result.
On the contrary, for instance, in $D=4$ space of Euclidian signature the
square of the self--dual combination of a gauge field--strength 
is no--zero and reproduces (up to a total derivative) the standard 
Maxwell Lagrangian, and no reasonable choice of its components is known
in these cases to construct actions analogous to \p{ss}.

We have seen that the method we used to get the action
breaks manifest Lorentz invariance,
however, beside the manifest spatial rotations the action 
\p{ss} is invariant under the following modified space--time 
transformations of $A^\alpha_i$ (in the gauge $A^\alpha_0=0$)
\begin{equation}\label{st}
\d A^\alpha_i=x^0v^k\partial_kA^\alpha_i+v^kx^k\partial_0A^\alpha_i+
v^kx^k{\cal L}^{\alpha\beta}{\cal F}^\beta_{0i},
\end{equation}
where the first two terms describe the ordinary Lorentz boosts along
a constant velocity $v^i$ and the third term vanishes 
on the mass shell since an additional local symmetry of the action 
\p{ss}
\begin{equation}\label{ad}
\d A^a_0=\varphi^\alpha(x)
\end{equation}
allows one to reduce the equations of motion 
\begin{equation}\label{em}
{{\d S}\over{\d A^\alpha_i}}=\e^{ijk}\partial_i{\cal F}^\alpha_{k0}=0
\end{equation}
to the duality condition
\begin{equation}\label{9}
{\cal F}^\alpha_{mn}={\cal L}^{\alpha\beta}F^\beta_{mn}-{1\over 
2}\epsilon_{mnlp}F^{lp\alpha}=0
\end{equation}
which, on the one hand, leads to the Maxwell equations
\begin{equation}\label{ma}
\partial_m{\cal F}^{mn\alpha}=\partial_mF^{mn\alpha}=0
\end{equation}
and, on the other hand, completely determines one of the gauge fields
through another one. For instance, using the relation
$$
{1\over 2}\epsilon_{mnlp}F^{2mn}=F^1_{mn}
$$
we can exclude $A^2_m(x)$ from the action \p{ss} and get the 
conventional Maxwell action.

One can admit that the action \p{ss}
arose as a result of some gauge fixing which specifies time direction
in a Lorentz covariant action \cite{pst}. 

The first step is to covariantize the self--dual part of the 
action \p{ss}. For people who are acquainted with harmonic techniques
served for similar covariantization purposes in supersymmetric theories
\cite{gikos} the first thing which comes into mind is
to introduce an 
auxiliary harmonic--like vector field 
\begin{equation}\label{harm}
l_m(x)\equiv {{u_m(x)}\over{\sqrt{-u_nu^n}}}, \qquad l_ml^m=-1,
\end{equation} 
and to write the action as follows:
\begin{equation}\label{11}
S_A=\int d^4x (-{1\over 8}F^\alpha_{mn}F^{\alpha mn}
+{1\over{4(-u_lu^l)}}u^m{\cal F}^\alpha_{mn}{\cal F}^{\alpha np}u_p).
\end{equation}
The field \p{harm} is harmonic in the sense that if supplemented
with space--like fields $l^i_m(x),~~l^i_ml^{jm}=\delta^{ij},~~l^ml_m^i=0$
(which do not enter the action) the set of the four vector fields
form a matrix of the Lorentz group $SO(1,3)$ and can be used to
contract Lorentz indices of other fields in a covariant way. 
For the analysis of properties
of the action it was proved more convenient
to work with the field $u(x)$ rather then with its normalized 
form $l(x)$, and at the same time to use harmonic properties of the 
latter.

The main problem is to find a local symmetry which would permit to 
choose a gauge where $u_m$ is a constant vector, in particular,
\begin{equation}\label{g}
u_m(x)=\d^0_m.
\end{equation}
Then the action \p{11} can reduce to \p{ss}. Note that a spatial
gauge, for instance $u_m(x)=\d^3_m$ is equally admissible and leads
to a non-covariant action which also produces the duality condition
\p{9}.  

The search for this symmetry turns out to be connected with another 
problem, namely, the problem of preserving a local symmetry under \p{ad}. 
In the covariant version this transformation should be replaced by
\begin{equation}\label{adc}
\d A^\alpha_m=u_m\varphi^\alpha.
\end{equation}
To keep this symmetry is important (as we have already seen) for getting 
the duality condition \p{9}.

To have the invariance under transformations \p{adc} one should add to 
the Lorentz invariant action \p{11} another term
\begin{equation}\label{B}
S_B=-\int d^4x\e^{mnpq}u_m\partial_nB_{pq},
\end{equation}
where $B_{mn}(x)$ is an antisymmetric tensor field. Then the variation 
of \p{11} under \p{adc} is canceled by the variation of \p{B} under
\begin{equation}\label{bv}
\d B_{mn}=-{{\varphi^\alpha}\over{u^2}}({\cal F}^\alpha_{mp}u^pu_n-
{\cal F}^\alpha_{np}u^pu_m).
\end{equation}
The equation of motion of $B_{mn}$ 
\begin{equation}\label{ue}
\partial_{[m}u_{n]}=0~~~\rightarrow~~~u_m(x)=\partial_ma(x)
\end{equation}
reads that $u_m(x)$ is the derivative of a scalar field.
Note also that \p{B} is invariant under
\begin{equation}\label{b}
\delta B_{mn}=\partial_{[m}b_{n]}(x).
\end{equation}

As in the case of the action \p{ss}, the local symmetry \p{adc} allows one to 
fix a gauge on the mass shell in such a way that the duality condition
\p{9} takes place. To arrive at eq. \p{9} harmonic techniques 
found to be rather useful. Let us sketch the derivation of \p{9}.

The equation of motion of $A^\alpha_m$ produced by \p{11} is
\begin{equation}\label{eq}
\epsilon^{mnpq}\partial_l(l_p{\cal F}^\alpha_q)=0,
\end{equation}
where ${\cal F}^\alpha_q\equiv {\cal F}^\alpha_{qp}l^p$ and $l_p$ is 
defined by \p{harm} and \p{ue}.

From \p{eq} it follows that
\begin{equation}\label{eq1}
l_{[p}{\cal F}^\alpha_{q]}=\partial_{[p}\Phi^\alpha_{q]}, \quad
{\cal F}^\alpha_{q}=l^p\partial_{p}\Phi^\alpha_{q}-
l^p\partial_{q}\Phi^\alpha_{p},
\end{equation}
where $\Phi^\alpha_q$ are two vector functions. Projecting \p{eq1}
onto harmonics $l^p_i$ $(i=1,2,3)$ orthogonal to $l_p$ we get
$$
l^p_{[i}l^q_{j]}\partial_{p}\Phi^\alpha_{q}=0,
$$
which implies that
\begin{equation}\label{eq2}
l^q_i\Phi^\alpha_{q}=l^q_i\partial_q\varphi^\alpha(x) ~~~\Rightarrow
~~~\Phi^\alpha_{q}=l_q\Phi^\alpha+\partial_q\varphi^\alpha(x).
\end{equation}
Substituting \p{eq2} into \p{eq1} and taking into account that the last
term in \p{eq2} can be neglected since \p{eq1} is invariant under gauge 
transformations $\Phi^\alpha_{q}\rightarrow\Phi^\alpha_{q}+
\partial_q\varphi^\alpha(x)$ we obtain
\begin{equation}\label{eq3}
{\cal F}^\alpha_q=\partial_q\Phi^\alpha(x)+l^p\partial_{p}
(l_q\Phi^\alpha).
\end{equation}
Now the transformation \p{adc} can be used to put in \p{eq3}
$\Phi^\alpha=0$ as a gauge fixing condition. Then we have
${\cal F}^\alpha_q=0$ which, because of the self--duality of
${\cal F}^\alpha_{pq}$, implies \p{9}.

Thus, we again remain with only one independent Maxwell field and get 
the duality between its electric and magnetic strength vector. In view of 
the vanishing condition for the self--dual strength tensor the equations
of motion of $u_m$ reduce to:
\begin{equation}\label{u}
{{\d(S_A+S_B)}\over{\d u_m}}=\e^{mnlp}\partial_nB_{lp}=0 ~~\rightarrow~~
B_{mn}=\partial_{[m}b_{n]},
\end{equation}
which means that $B_{mn}$ is completely auxiliary and can be eliminated 
by use of the corresponding local transformations \p{b}.

The only thing which has remained to show is that $u_m$ itself does not 
carry physical degrees of freedom and can be gauge fixed to 
$u_m=\d^0_m$. For this we have to find a corresponding local symmetry.
And here an analogy of the antisymmetric field $B_{mn}$ 
with the `notoph' of Ogievetsky and Polubarinov \cite{o} helps us
to get the corresponding symmetry transformations.
 
The form of the action \p{B} containing $B_{mn}$ reminds a 
term which one encounters in a dual formulation of a pseudoscalar 
(`axion') field 
as an antisymmetric notoph field 
(see, for instance, \cite{o,bo}) 
\begin{equation}\label{1}
S=\int d^4x \left(
-{1\over 2}(\partial_m a(x)-u_m(x))(\partial^m a(x)-u^m(x))
-\e^{pqmn}u_p\partial_qB_{mn} \right).
\end{equation}

The action \p{1} is invariant under local Peccei--Quinn transformations
\begin{equation}\label{pct}
\d a(x)=\varphi(x),\qquad \d u_m=\partial_m\varphi(x),
\end{equation}
($u_m$ being the corresponding gauge field) and produces dual 
versions of the axion action\footnote{We denoted the scalar field in 
\p{ue} with the same letter
$a(x)$ as the axion field to point to their formal ``generic 
roots".}:
$$
L=-{1\over 2}\partial_ma(x)\partial^ma(x),
$$
\begin{equation}\label{a}
L={1\over{3!}}\partial_{[m}B_{np]}\partial^{[m}B^{np]}.
\end{equation}
The duality relation between the pseudoscalar field $a(x)$ and the 
antisymmetric tensor field $B_{mn}$
\begin{equation}\label{5}
\partial_la(x)=\epsilon_{lmnp}\partial^mB^{np}
\end{equation}
is a consequence of the equations of motion of $u_m$ obtained from
\p{1}.

Now one can assume that the action \p{11}+\p{B} is also invariant under
the transformations \p{pct}. 
This is indeed the case provided $A^\alpha_m$ and $B_{mn}$ transform as 
follows
\begin{equation}\label{12}
\d A^\alpha_m=
{{\varphi(x)}\over{u^2}}{\cal L}^{\alpha\beta}{\cal F}^\beta_{mn}u^n,
\qquad
\d B_{mn}={{\varphi(x)}\over{(u^2)^2}}{\cal F}^{\alpha r}_{m}
u_r{\cal F}^{\beta s}_nu_s{\cal L}^{\alpha\beta}.
\end{equation}
Then, taking into account \p{ue} 
and requiring that $u^2\not= 0$ (to escape singularities), we can use 
the local transformation \p{pct} to put $u_m=\d^0_m$. 
In this gauge the manifestly 
Lorentz invariant duality--symmetric action
\begin{equation}\label{lo}
S=\int d^4x (-{1\over 8}F^\alpha_{mn}F^{\alpha mn}
-{1\over{4(u_lu^l)}}u^m{\cal F}^\alpha_{mn}{\cal F}^{\alpha np}u_p
-\epsilon^{mnpq}u_m\partial_nB_{pq})
\end{equation}
reduces to \p{ss}, and the local transformations of 
$A_m^\alpha$ \p{12} (with $\varphi(x)=x^iv^i$) are combined with the 
corresponding Lorentz transformations and produce the modified 
space--time symmetry \p{st} of the action \p{ss}.

One can reduce the number of the auxiliary fields 
in the action \p{lo} to one scalar field by substituting into \p{lo} the 
solution of the equation of motion \p{ue}. The resulting minimal form of
the covariant duality symmetric action 
\begin{equation}\label{ax}
S=\int d^4x (-{1\over 8}F^\alpha_{mn}F^{\alpha mn}
-{1\over{4(\partial_la\partial^la)}}
\partial^ma(x){\cal F}^\alpha_{mn}{\cal F}^{\alpha 
np}\partial_pa(x)),
\end{equation}
which remains the same in all space--time dimensions,
has been used for the description
of chiral bosons in various theoretical models (see \cite{pst1,bbs,alst}
and references therein).

We have thus obtained a covariant Lagrangian formulation of Maxwell 
theory which is also invariant under electric--magnetic duality. The 
action was shown to produce in the temporal gauge \p{g} 
the duality--symmetric action \p{ss} of \cite{d,ss}. 

As has been mentioned in the introduction the first duality--symmetric 
action for the Maxwell fields was constructed by Zwanziger \cite{z}
and that it differs from \p{ss} in the sign of the second term.
This difference leads to essentially different symmetry properties
of the Zwanziger action \cite{mps}.
However, since both actions describe one and the same physical model 
there should
be a relationship between them. This relation is established through a 
duality transform of the auxiliary scalar field $a(x)$ into the
auxiliary `notoph' field $B_{mn}$ \cite{mps}. For this consider \p{lo}
as a master action which produces different dual actions depending on
which auxiliary fields are integrated out. The action \p{ax} is one
of these dual actions. Another one is obtained by varying \p{lo} with 
respect to $u_m$, which gives an expression for the dual field 
strength $v^m=\epsilon^{mnpq}\partial_nB_{pq}$ 
in terms of $u_m$ and the Maxwell field strengths, solving this 
expression for the vector field $u_m$ in terms of $v_m$ and substituting
the result back into the action \p{lo}. The action now contains only
the dual field strength $v_m$ of the auxiliary field $B_{mn}$ and has
the form
\begin{equation}\label{bx}
S=\int d^4x (-{1\over 8}F^\alpha_{mn}F^{\alpha mn}
+{1\over{4(v_lv^l)}}v^m{\cal F}^\alpha_{mn}{\cal F}^{\alpha np}v_p),
\end{equation}  
the sign of the second term being changed with respect to the analogous
term in \p{lo} and \p{ax}.
It can be shown that the action \p{bx} is invariant under
local transformations which allow one to fix $v_m(x)$ to be a constant
time--like or space--like vector upon which \p{bx} reduces to the
Zwanziger action (see \cite{mps} for details).

In conclusion we have shown how the covariant Lagrangian approach to the 
description of duality--symmetric fields unifies different 
non--covariant formulations. This approach is also related to the 
infinite field approach \cite{inf} being a consistent truncation of the 
latter (see the last ref. in \cite{pst}).

From the action \p{ax} one can {\it formally} obtain the Siegel action 
\cite{si} by replacing ${{\partial_ma(x)\partial_pa(x)}
\over{(\partial_la\partial^la)}}$ with a Lagrange multiplier field 
$\Lambda_{mp}(x)$. But this relation is only a formal one since 
``hiding" derivatives of fields in other fields is not an innocent 
trick. The properties of the two actions are very different, a main 
difference being that the duality condition \p{9} is obtained from the 
actions \p{ss}, \p{lo}--\p{bx} as a consequence of equations of motion 
of the gauge fields $A^\alpha_m(x)$, while in the Siegel formulation it 
arises as a ``square root" of the constraint produced by the 
Lagrange multiplier $\Lambda_{mn}$ 
equation of motion. This results in a different structure of Hamiltonian
constraints and, as a consequence, leads to different ways of quantizing
chiral bosons. We also note that for more complicated cases of 
self--interacting chiral bosons, such as the M--theory five--brane 
\cite{pst1} an effective action in the Siegel form is not know yet.

\bigskip
{\bf Acknowledgements}. This work was partially supported by the 
European Commission TMR Programme ERBFMPX-CT96-0045 to which the authors 
are associated, and by the INTAS Grant 96-308.

\end{document}